\documentclass[12pt,twoside]{article}

\usepackage{a4wide,cite}
\usepackage{amsmath,amssymb}
\usepackage{epsfig,rotating}
\usepackage{axodraw}
\usepackage{color}

\numberwithin{equation}{section}
\allowdisplaybreaks[4]

\def \be{\begin{equation}}
\def \ee{\end{equation}}
\newcommand{\bea}{\begin{eqnarray}}
\newcommand{\eea}{\end{eqnarray}}

\begin{document}
\begin{titlepage}

\vskip 2cm

\centerline{\large {\bf Tree-level contribution to} {\boldmath $\bar B \to X_d \gamma$}
{\bf using fragmentation functions}}
\vskip 1cm

\begin{center}
  {\bf H.M.~Asatrian$^a$, C.~Greub$^b$}\\[1mm]
  {$^a$\sl Yerevan Physics Institute, 0036 Yerevan, Armenia}\\[1mm]
  {$^b$\sl Albert Einstein Center for Fundamental Physics, Institute 
  for Theoretical Physics,\\
 Univ.~of Bern, CH-3012 Bern, Switzerland}\\[1mm]
\end{center}
\medskip

\vskip 1cm

\begin{abstract}
\noindent 
We evaluate the most important tree-level contributions connected with
the $b \to u\bar u d\gamma$ transition 
to the inclusive radiative decay $\bar B \to X_d \gamma$
using fragmentation functions. 
In this framework
the singularities arising from collinear photon emission 
from the light quarks  ($u, \bar{u}$ and $d$)  
can be absorbed into the (bare) quark-to-photon 
fragmentation function.
We use as input the fragmentation function 
extracted by the ALEPH group from the two-jet cross section measured at LEP, 
where one of the jets is required to contain a photon.
To get the quark-to-photon fragmentation
function at the fragmentation scale $\mu_F \sim m_b$, we use
the evolution equation,
which we solve numerically. We then calculate the (integrated) photon
energy spectrum for $b \to u \bar{u} d \gamma$ related to the
operators $P^u_{1,2}$.
For comparison, we also give the corresponding results when using
nonzero (constituent) masses for the light quarks. 

\end{abstract}
\end{titlepage}

\section{Introduction}
Rare B-meson decays  are known to be a unique source of indirect 
information about physics at scales of several hundred GeV. 
To make a rigorous comparison between experiment and theory, one needs 
precise theoretical predictions for them.

\noindent  The first estimate of the $\bar{B}\to X_s\gamma$ branching
ratio within the Standard 
Model (SM) at the next-to-next-to-leading logarithmic (NNLL) level was 
published some years ago 
\cite{Misiak:2006zs}. The branching ratio of the $\bar{B}\to X_s\gamma$ decay
has been measured at the B factories \cite{Chen:2001fj}. 
Both theory and experiment have acquired a precision 
of a few percent.

\noindent  Another interesting process is $\bar B \to X_d \gamma$. 
In \cite{Ali:1998rr} its decay rate 
was calculated in next-to-leading logarithmic (NLL) order
(earlier in \cite{Ali:1992nv} a partial NLL result was found). 
Since then, not much theoretical work was 
done on  $\bar B \to X_d \gamma$, 
because the corresponding measurement seemed very difficult. 
A few years ago, however, the BABAR collaboration managed to measure 
this branching ratio \cite{Aubert:2008av,delAmoSanchez:2010ae},
yielding the CP-averaged result
$Br[\bar{B}\to X_d \gamma]_{E_{\gamma}>1.6 \, \rm{GeV}}=(1.41\pm0.57)\cdot 10^{-5}$
\cite{Wang:2011sn,Crivellin:2011ba}.   
Extrapolation factors \cite{Amhis:2012bh} 
were used to get from the experimental data to this result
which, as indicated by the notation, corresponds to a photon energy cut
of 1.6 GeV. 

Non-perturbative contributions related to $u$-quark loops
are not CKM suppressed in $\bar{B} \to X_d \gamma$ (unlike in $\bar{B} \to X_s
\gamma$) and therefore potentially limit the theoretical precision.
It was, however, realized recently that most of these uncertainties
drop out in the CP-averaged branching ratio 
\cite{Benzke:2010js,Hurth:2010tk}.
This
implies that the SM predicition for $\bar{B} \to X_d \gamma$ can in
principle be calculated with similiar accuracy as $\bar{B} \to X_s \gamma$. 
To fully exploit the information from $\bar{B} \to X_d \gamma$,
we plan to derive in the near future a NNLL prediction of its CP averaged
branching ratio.  

When going to this precision with the $b \to d \gamma$ transition, one has
to take into account also the contributions from the tree-level transitions 
$b\to u \bar{u} d \gamma$, which lead to a non-negligible contribution
to the inclusive photon energy spectrum in $\bar{B} \to X_d \gamma$ in the
considered photon energy window. 
Analogous tree-level contributions were considered some time ago
for the full-inclusive photon energy spectrum of $B$ decays
\cite{Ali:1992qs} and more recently for the $\bar B \to X_s \gamma$ decay
in \cite{Kaminski:2012eb}. In the calculation of the Feynman diagrams singularites arise
which are related to collinear photon emission from one of the light
quarks $q$ ($q=u, \bar{u}$ or $d$), leading to single
logarithms of the form $\ln(m_b^2/m_q^2)$. In the references just
mentioned, the current quark masses $m_q$ are replaced by a common
constituent mass, providing an effective treatment of the collinear
configurations.

In the present paper we evaluate the mentioned tree-level transitions to
$\bar{B} \to X_d \gamma$ associated with the operators $P^u_1$ and $P^u_2$ (see
eq. (\ref{p1p2})), using the fragmentation function approach.
In this framework
the singularities arising from collinear photon emission 
from the light quarks  ($u, \bar{u}$ and $d$)  
can be absorbed into the (bare) quark-to-photon 
fragmentation function. 

Using this framework, one often starts with a non-perturbative initial condition for
the fragmentation function at a low scale $\mu_F=\mu$ ($\mu$ of order $\Lambda_{\rm{QCD}}$) and then evolves
it up to the scale $\mu_F=m_b$, using the inhomogeneous evolution
equation. When using a zero initial condition, the (LL) evolution equation 
resums the terms $\left[\alpha_s \log(m_b^2/\mu^2)\right]^n$. This resummation 
was done in \cite{Kapustin:1995fk} in connection with
the $P_8$ contribution to $\bar{B} \to X_s \gamma$. It was found that the
effect of resummation suppresses the corresponging single logarithm
present in  lowest order perturbation theory for $E_\gamma>1.6$ GeV,
which is the relavant range in our application. In
\cite{Ferroglia:2010xe}, dealing also with the 
$P_8$ contribution, a model based on vector meson dominance was
used for the initial condition of the quark-to-photon fragmentation
function \cite{Bourhis:1997yu}.
As in our application the logarithmic term (at lowest order) is
numerically not much larger than the non-logarithmic one, and because
the two contributions have a different sign, we have some doubts to get
reasonable predictions along these lines. This feature related to size
and sign can also be seen in Table II of ref. \cite{Kaminski:2012eb}.
 
As we will describe in detail in Sec. 2, we use as input the fragmentation function 
extracted by the ALEPH group from the two-jet cross section measured at LEP, 
where one of the jets is required to contain a photon \cite{Buskulic:1995au}. The
theoretical framework needed for the extraction of the fragmentation
function at the scale $m_Z$ is described in
\cite{Glover:1993xc}. Radiative corrections are considered
in \cite{GehrmannDeRidder:1997wx,GehrmannDeRidder:1997gf,
GehrmannDeRidder:1998ba}.
To get the quark-to-photon fragmentation
function at the fragmentation scale $\mu_F \sim m_b$, we use
the (inhomogeneous) evolution equation \cite{Kunszt:1992np,GehrmannDeRidder:1998ba},
which we solve numerically. We then calculate the (integrated) photon
energy spectrum for $b \to u \bar{u} d \gamma$ related to $P^u_{1,2}$.
Then, for comparison, we also give the corresponding results when using nonzero (constituent)
masses for light quarks. 

\noindent As just mentioned, we  consider  in this paper the
contributions to $\bar{B} \to X_d \gamma$
originating from  the
four-quark operators $P^u_1$ and $P^u_2$
\be
\label{p1p2}
P_1^u=\left (\bar{d}_L\gamma_{\mu}T^au_L\right)
\left (\bar{u}_L\gamma_{\mu}T^ab_L\right),\,\,\,
P_2^u=\left (\bar{d}_L\gamma_{\mu}u_L\right)
\left (\bar{u}_L\gamma_{\mu}b_L\right)
\ee
appearing in the effective weak Hamiltonian for the process 
$\bar B \to X_d \gamma$
\be 
{\cal H}_{\rm weak} = 
 - \frac{4 G_F}{\sqrt{2}} \left[\xi_t \sum_{i=1}^{8} C_i P_i
                            +   \xi_u \sum_{i=1}^{2} C_i (P_i - P^u_i) \right],
\label{lang}
\ee
  \noindent  with $\xi_t=V_{tb} V_{td}^*$, $\xi_u=V_{ub}
  V_{ud}^*$. The other operators can be found in 
\cite{Kaminski:2012eb}. 
\noindent While $\xi_t$ and $\xi_u$ are numerically of the same order,
the Wilson coefficients of the four quark operators
$P_{i}$, $(i=3,..,6)$  are smaller than those of $P^u_{1,2}$. We
therefore only use the latter for our estimate.

\noindent The paper is organized as follows: 
In Sec.~2 we provide some details
about our calculation and present our analytic result for the
$P^u_{1,2}$ contribution
to the (integrated) photon energy spectrum, 
using the  fragmentation function approach. 
In Sec.~3 we give the corresponding results when using nonzero
(constituent) masses for the light quarks. 
Finally, we present our conclusions in Sec.~4.

\section{Fragmentation function approach to estimate the $P^u_{1,2}$ 
contributions}
\label{fragsection}
In this section we work out the upper end of the photon energy
spectrum (typically above 1.6 GeV) resulting from the tree-level
transitions $b \to  u \bar{u} d \gamma$ associated with the operators 
$P^u_1$ and $P^u_2$. 
\begin{figure}[h]
\begin{center}
\includegraphics[width=9cm]{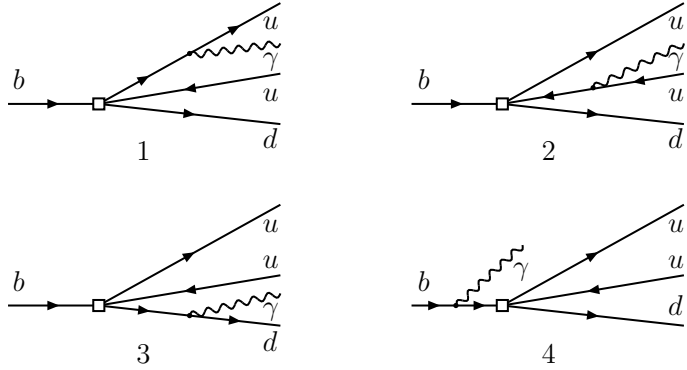}
\caption{Tree level Feyman diagrams corresponding to the transition 
$b\to u\bar{u}d\gamma$. The white square symbolizes the operators $P^u_{1,2}$.
\label{feyn}}
\end{center}
\end{figure}
The corresponding Feynman diagrams are shown in Fig.~\ref{feyn}. The 
kinematical configurations in which the photon is
emitted (almost) collinear with $d$, $u$ or $\bar{u}$ lead to
uncancelled singularities. This signals that there is another,
non-perturbative contribution. Indeed, the photon energy spectrum
$d\Gamma(b \to u \bar{u} d \gamma)/dE_\gamma$ (in short-hand notation
$d\Gamma/dE_\gamma$ in the following) can be
written as
\be
\label{width1}
\frac{d\Gamma}{dE_{\gamma}}=\frac{d\hat{\Gamma}}{dE_{\gamma}}+
\sum_{q=u,\bar{u},d}\int \frac{d\hat{\Gamma}}{dE_{q}} dE_q \, dz
\delta\left(E_{\gamma}-zE_q\right) \, D_{q\to\gamma}(z),
\ee
where the first term comes from the diagrams in 
Fig.~\ref{feyn}, while the 
second term involves the inclusive fragmentation functions
$D_{q\to\gamma}(z)$ for the partonic transitions $q \to q \gamma$, as
well as the 
energy spectrum $\frac{d\hat{\Gamma}}{dE_{q}}$ of the parton $q$
in the process $b\to u \bar{u}d$. The argument $z$ of the
fragmentation function $D_{q\to\gamma}$ denotes the energy fraction of the photon
relative to the energy of the parent parton $q$.

To make the paper self-contained, we briefly show in subsection \ref{analres}
that the mentioned
collinear singularities in the first term of eq. (\ref{width1}),
which we regulate dimensionally ($d=4-2\varepsilon$), 
can be factorized into the fragmentation
function. Doing so, we closely follow  
the formalism in
ref. \cite{Glover:1993xc}.
We then give the result for $d\Gamma(b \to u \bar{u} d
\gamma)/dE_\gamma$ in a way
where the singularites are manifestly cancelled. Subsection \ref{fragfunction}
then deals with the details of the fragmentation function
$D(z,\mu_F)$. Finally, in subsection \ref{numres} the numerical results
for the (integrated) photon energy spectrum are given.
\subsection{Analytic results for $d\Gamma(b \to u  \bar{u}d\gamma)/dE_\gamma$}
\label{analres}
For the following it is useful to decompose 
the first term in eq. (\ref{width1}) according to
\be
\label{width2}
\frac{d\hat{\Gamma}}{dE_{\gamma}}=\frac{d\hat{\Gamma}^{uu}}{dE_{\gamma}}+
\frac{d\hat{\Gamma}^{\bar{u}\bar{u}}}{dE_{\gamma}}+
\frac{d\hat{\Gamma}^{dd}}{dE_{\gamma}}+
\frac{d\hat{\Gamma}^{bb}}{dE_{\gamma}}+
\frac{d\hat{\Gamma}^{int}}{dE_{\gamma}},
\ee
where $\frac{d\hat{\Gamma}^{uu}}{dE_{\gamma}}$ corresponds to the 
self-interference 
of diagram 1 in Fig.~\ref{feyn}, i.e. 
with photon emission from the $u$ quark,
$\frac{d\hat{\Gamma}^{\bar{u}\bar{u}}}{dE_{\gamma}}$ corresponds to
the self-interference of diagram 2, etc.
When summing over the transverse polarizations of the photon, we 
used the general expression
\be
\label{pol1}
\sum_{r=1}^2\epsilon_{r,\mu}\epsilon_{r,\nu}=-g_{\mu\nu}+
\frac{t_{\mu}k_{\nu}+k_{\mu}t_{\nu}}{k\cdot t}-\frac{t^2k_{\mu}k_{\nu}}{(k\cdot t)^2},
\ee
where the polarization vectors $\epsilon_r$, the momentum $k$ of the
photon and the arbitrary vector $t$ are chosen such that
\be
\label{pol2}
\epsilon_r\cdot k = 0,\qquad \epsilon_r\cdot \epsilon_{r'} =
-\delta_{rr'},\qquad
\epsilon_r\cdot t = 0 \qquad (r,r' = 1,2) \, .
\ee
The vector $t$ is then identified with the momentum of the $b$-quark, i.e. 
$t=p_b=m_b(1,0,0,0)$. It turns out that in this setup the  
collinear singularities are contained in the contributions
$\frac{d\hat{\Gamma}^{uu}}{dE_{\gamma}}$,
$\frac{d\hat{\Gamma}^{\bar{u}\bar{u}}}{dE_{\gamma}}$ and
$\frac{d\hat{\Gamma}^{dd}}{dE_{\gamma}}$, whereas the sum of the other two terms
define the non-singular (ns) contribution
$\frac{d\Gamma^{ns}}{dE_{\gamma}}$,
\be
\label{partns}
\frac{d\Gamma^{ns}}{dE_{\gamma}} =
\frac{d\hat{\Gamma}^{bb}}{dE_{\gamma}} +
\frac{d\hat{\Gamma}^{int}}{dE_{\gamma}} \, .
\ee
As a consequence of this specific singularity structure, it is useful to
 consider the combinations
$\frac{d\Gamma^{qq}}{dE_{\gamma}}$ (with $q=u,\bar{u},d$) defined as
\be
\label{fr1}
\frac{d\Gamma^{qq}}{dE_{\gamma}}=\frac{d\hat{\Gamma}^{qq}}{dE_{\gamma}}+
\int \frac{d\hat{\Gamma}}{dE_{q}} dE_q \, dz \,
\delta\left(E_{\gamma}-z \, E_q\right) \, D_{q\to\gamma}(z)
 \ee 
To be concrete, we work out in the following
$d\Gamma^{uu}/dE_{\gamma}$ (the other two quantities can then be obtained
by obvious replacements).
Defining 
$s=(p_u+p_\gamma)^2$, we can split the perturbative part
$d\hat{\Gamma}^{uu}/dE_\gamma$ according to
\be
\label{fr2}
\frac{d\hat{\Gamma}^{uu}}{dE_{\gamma}}=\frac{d\hat{\Gamma}^{uu,res}}{dE_{\gamma}}
 +\frac{d\hat{\Gamma}^{uu,coll}}{dE_{\gamma}}
\ee
where the resolved (collinear) contribution corresponds to 
$s>s_{min}$ ($s<s_{min}$).
Needless to say, the double differential decay width 
$\frac{d\hat{\Gamma}^{uu}}{ds \, dE_{\gamma}}$ needs to be worked out to obtain this splitting.
The resolved part is finite for each $s_{min}>0$. If $s_{min}$ is sufficiently small,
the collinear piece can be worked out using the collinear approximation of
the matrix element and the phase space, leading to
\bea
\label{fr3}
\frac{d\hat{\Gamma}^{uu,coll}}{dE_{\gamma}}&=&
\int \frac{d\hat{\Gamma}}{dE_{u}} dE_u \, dz \,
\delta\left(E_{\gamma}-z \, E_u\right)
\left[
-\frac{1}{\varepsilon}\left (\frac{4 \pi \mu^2}{s_{min}}\right)^{\varepsilon}   
\frac{[z(1-z)]^{-\varepsilon}}{\Gamma(1-\varepsilon)} \,
\frac{\alpha e_u^2}{2 \pi} \, P(z) \right],
\eea
where the $d$-dimensional splitting function $P(z)$ reads
\be
\label{fr4}
P(z)=\frac{1+(1-z)^2-\varepsilon z^2}{z} \, . 
\ee
From the explicit structure of $\frac{d\hat{\Gamma}^{uu,coll}}{dE_{\gamma}}$
we see that the $1/\varepsilon$ pole can be factorized into the 
bare fragmentation function $D_{u\to\gamma }(z)$ at the fragmentation scale
$\mu_F$ such that in the $\overline{\rm{MS}}$ scheme 
\be
\label{fr5}
D_{u\to\gamma }(z)=D_{u\to\gamma }(z,\mu_F)+
\frac{1}{\varepsilon}\left (\frac{4 \pi \mu^2}{\mu_F^2}\right
)^{\varepsilon} \,
  \frac{1}{\Gamma(1-\varepsilon)} \, \frac{\alpha e_u^2}{2 \pi} \, P^{(0)}(z), 
\ee
where the four-dimensional splitting function $P^{(0)}(z)$ reads 
\be
\label{splittingfour}
P^{(0)}(z)=\frac{1+(1-z)^2}{z} \, .
\ee
The result for $\frac{d\Gamma^{uu}}{dE_{\gamma}}$ is then
\bea
\label{fr6}
\frac{d\Gamma^{uu}}{dE_{\gamma}}&=&\frac{d\hat{\Gamma}^{uu,res}}{dE_{\gamma}}+
\nonumber\\
&&\int \frac{d\hat{\Gamma}}{dE_{u}} dE_u \, dz \,
\delta\left(E_{\gamma}-zE_u\right) \,
\frac{\alpha e_u^2}{2 \pi} \left[ P^{(0)}(z)
\ln\left (\frac{s_{min}z (1-z)}{\mu_F^2}\right )+z\right] +
\nonumber\\
&&\int \frac{d\hat{\Gamma}}{dE_{u}} dE_u \,dz \,
\delta\left(E_{\gamma}-zE_u\right) \,
D_{u\to \gamma}\left (z,\mu_F \right) \, .
\eea
The treatment of the collinear phase space becomes exact when we take the limit
$s_{min}\to 0$. 
After  working out the integral in the second term, we get 
in this limit (using $e_{\gamma}=E_{\gamma}/m_b$) 
\bea
\label{fr7}
\frac{d\Gamma^{uu}}{dE_{\gamma}}&=&\frac{\alpha e_u^2}{2 \pi} \, 
\frac{G_F^2m_b^4|\xi_u|^2}{288 \pi^3} \,
\frac{\left(9 \, C_2^2 + 2 \, C_1^2\right)}{3} \, \frac{1}{e_{\gamma}} \, 
\left[ 2 \left(24 e_{\gamma}^2-10 e_{\gamma}+3\right)
   \ln (e_{\gamma})\right. \nonumber\\
&&+(1-2 e_{\gamma}) \left(16 e_{\gamma}^3-16 e_{\gamma}^2+4
   e_{\gamma}-3\right) \ln
   \left(\mu_F^2/m_b^2\right)\nonumber\\
&&
-(1-2 e_{\gamma}) \left(16 e_{\gamma}^3-16 e_{\gamma}^2+4
   e_{\gamma}-3\right)  \ln (1-2
   e_{\gamma})\\ 
   &&\left.-16 e_{\gamma}^3(3+e_{\gamma})+e_{\gamma}^2
   (38+48 \ln (2))+e_{\gamma}(1-20 \ln (2))-3+6\ln (2)\right]\nonumber\\        
&&+\int \frac{d\hat{\Gamma}}{dE_{u}} dE_u \, dz \,
   \delta\left(E_{\gamma}-zE_u\right) \, 
D_{u\to \gamma}\left (z,\mu_F \right),\nonumber
\eea
where
\be
\frac{d\hat{\Gamma}}{dE_{u}}=G_F^2 \, m_b|\xi_u|^2 \, \frac{\left( 9 \, C_2^2 +
  2 \, C_1^2 \right)}{3} \,
\frac{E_u^2(3 m_b -4 \, E_u)}{12\pi^3} \, .
\ee
We notice that from this procedure we could easily read-off the renormalization 
equation (\ref{fr5}). Assuming that this universal equation is known already (which is of 
course the case), we could have obtained the final result (\ref{fr7})  
without seperating the phase  space into a resolved and a collinear region, which 
has the advantage that the double differential decay width (in
$E_\gamma$ and $s$) 
is not needed. Indeed, as a check, we followed this procedure and got the same
result.

\noindent In an analogous way we obtain  
$\frac{d\Gamma^{\bar{u}\bar{u}}}{dE_{\gamma}}$:
\bea
\label{fr8}
\frac{d\Gamma^{\bar{u}\bar{u}}}{dE_{\gamma}}&=&\frac{\alpha e_u^2}{2
  \pi} \,
\frac{G_F^2m_b^4|\xi_u|^2}{96 \pi^3} \, 
\frac{\left( 9 \, C_2^2+2 \, C_1^2 \right)}{3} \, \frac{1}{e_{\gamma}} \, 
\left[2 \left(12 e_{\gamma}^2-4 e_{\gamma}+1\right)
   \ln (e_{\gamma})\right. \nonumber\\
&&-(1-2 e_{\gamma})^2 \left(8
   e_{\gamma}^2+1\right) \ln
   \left(\mu_F^2/m_b^2\right)\nonumber\\
&&
+(1-2 e_{\gamma})^2 \left(8
   e_{\gamma}^2+1\right) \ln (1-2 e_{\gamma})\\ 
   &&\left.-4e_{\gamma}^3(5+6e_{\gamma})+6 e_{\gamma}^2
   (3+4 \ln (2))+e_{\gamma}(1-8  \ln(2))-1+2 \ln (2)\right]\nonumber\\        
&&+\int \frac{d\hat{\Gamma}}{dE_{\bar{u}}} dE_{\bar{u}} \, dz \, \delta\left(E_{\gamma}-
zE_{\bar{u}}\right) \, 
D_{\bar{u}\to \gamma}\left (z,\mu_F \right)\nonumber \, ,
\eea
where
\be
\frac{d\hat{\Gamma}}{dE_{\bar{u}}}=G_F^2m_b|\xi_u|^2 \,
\frac{\left(9 \, C_2^2 + 2 \, C_1^2 \right)}{3} \,
\frac{E_{\bar{u}}^2(m_b-2 E_{\bar{u}})}{2\pi^3} \, .
\ee
We note that in our approximation the fragmentation functions
$D_{q \to \gamma}(z,\mu_F)$ are the same for $q=u,\bar{u},d$ up to
obvious charge factors.

\noindent $\frac{d\Gamma^{dd}}{dE_{\gamma}}$ is easily obtained as it is related to 
$\frac{d\Gamma^{uu}}{dE_{\gamma}}$ through a Fierz identity. One obtains
\be
\label{fr9}
\frac{d\Gamma^{dd}}{dE_{\gamma}}=
\frac{e_d^2}{e_u^2} \, \frac{d\Gamma^{uu}}{dE_{\gamma}} \, .
\ee
Finally, the explicit expression for the non-singular part $d\Gamma^{ns}/dE_\gamma$, defined in
eq. (\ref{partns}), reads
\be
\label{fr10}
\frac{d\Gamma^{ns}}{dE_{\gamma}}=
\frac{\alpha }{2 \pi} \,
\frac{G_F^2m_b^4|\xi_u|^2}{2592 \, \pi^3} \,
\frac{\left(9 \, C_2^2+ 2 \, C_1^2\right)}{3} \, \frac{\left( 1- 2 \,
  e_{\gamma} \right)}{e_{\gamma}} \, 
\left(14 e_{\gamma}^3+39 e_{\gamma}^2-27 e_{\gamma}+14 \right) \, .
\ee
To summarize this subsection, the photon energy spectrum $d\Gamma(b \to u \bar{u} d \gamma)/dE_\gamma$ can be written as
\be
\label{summar}
\frac{d\Gamma(b \to u \bar{u} d \gamma)}{dE_\gamma} = 
\frac{d\Gamma^{uu}}{dE_\gamma} +
\frac{d\Gamma^{\bar{u}\bar{u}}}{dE_\gamma} +
\frac{d\Gamma^{dd}}{dE_\gamma} +
\frac{d\Gamma^{ns}}{dE_\gamma} \, ,
\ee
where the individual terms are given in eqs. (\ref{fr7}), (\ref{fr8}),
(\ref{fr9}) and (\ref{fr10}). 
\subsection{Parametrization of the fragmentation function}
\label{fragfunction}
To obtain a numerical prediction for $d\Gamma(b \to u \bar{u} d
\gamma)/dE_\gamma$ we need the non-perturbative fragmentation function $D_{q
  \to \gamma}(z,\mu_F)$ at a fragmentation scale $\mu_F$ which is
typically of order $ m_b$.
As we are not aware that this function has been directly extracted
at the factorization scale $\mu_F \sim m_b$, i.e. from $B$-decays,
we use as an input the ALEPH measurement of the fragmentation
function obtained for $\mu_F=m_Z$ \cite{Buskulic:1995au}. 

We should stress here that 
this measurement was induced by a theoretical paper of Glover and
Morgan \cite{Glover:1993xc} who suggested to extract this function
from the normalized 2-jet cross section
\be
\label{twojet}
\frac{1}{\sigma_{\rm had}} \, \frac{d\sigma(2-jets)}{d z_\gamma}
\ee
in $e^+e^-$ collisions at the $Z$-pole. To be more precise, the photon
is understood to be within one of the two jets ($E_\gamma>5$ GeV),
carrying at least $70\%$ of the total energy of the jet. The
fractional energy, $z_\gamma$, of such a photon within a jet is
defined as
\be
z_\gamma = \frac{E_\gamma}{E_\gamma+E_{had}} \, ,
\ee
where $E_{had}$ is the energy of all accompanying hadrons in the
``photon jet''.

For the extraction of the fragmentation function
$D_{q\to\gamma}(z,\mu_F)$ at $\mu_F=m_Z$ the experimental paper
\cite{Buskulic:1995au} follows exactly the procedure described in
full detail in the theoretical work \cite{Glover:1993xc}. As a result,
the ALEPH experiment 
found that the simple ansatz  
\be
\label{fragansatz}
D_{q\to \gamma}\left (z,m_Z \right)= \frac{\alpha \, e_q^2}{2\pi} \left[ P^{(0)}(z)
\ln \left( \frac{m_Z^2}{\mu_0^2(1-z)^2} \right)  
 -1 - \ln \left( \frac{m_Z^2}{2\mu_0^2}\right) \right] \, ,
\ee
which contains a single parameter ($\mu_0$), leads to a reasonable 
description of the 2-jet cross section
(\ref{twojet}) for  
\be
\label{rangetwojet}
\mu_0= \left( 0.14^{+0.21+0.22}_{-0.08-0.04} \right) \, \mbox{GeV}  \, .
\ee
This analysis was done for values of $z_\gamma>0.7$, which means that
the measurement of the fragmentation function is  restricted to values
$z>0.7$.

In our problem we need the fragmentation functions $D_{q
  \to\gamma}(z,\mu_F)$ at the low scale $\mu_b$, where $\mu_b$ is of order $m_b$. To this end we
solve the corresponding evolution equation. In leading logarithmic
precision (w.r.t. QCD) $D_{q\to \gamma}(z,\mu_F)$ satisfied the
inhomogeneous integro-differential 
equation (see e.g. \cite{Kunszt:1992np,GehrmannDeRidder:1998ba})
\begin{figure}[h]
\begin{center}
\includegraphics[width=9cm]{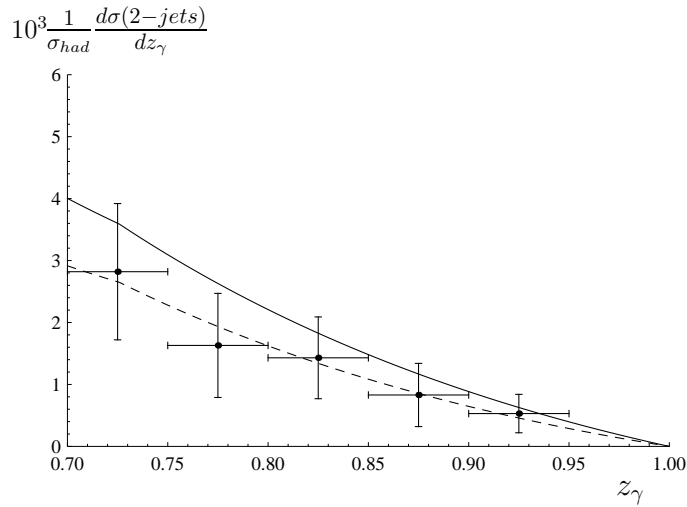}
\caption{Two-jet cross section $\frac{1}{\sigma_{had}} \,
  \frac{d\sigma(2-jets)}{dz_\gamma}$ for $y_{\rm{cut}}=0.06$. The dots
  correspond to the ALEPH measurements \cite{Buskulic:1995au} (see
  there Fig.~5 and Table 2), while
  the dashed line shows the theory prediction when using the central value
  $\mu_0=0.14$ GeV in the parametrization (\ref{fragansatz}) of the
  fragmentation function $D_{q \to \gamma}(z,m_Z)$. The solid line
  corresponds to $\mu_0=0.02$ GeV (see text).
\label{fit}}
\end{center}
\end{figure}
\be
\label{llevolution}
\mu_F \, \frac{\partial D_{q\to \gamma}\left (z,\mu_F \right)}{\partial \mu_F}=
\frac{\alpha \, e_q^2}{\pi} P^{(0)}(z)+
\frac{\alpha_s(\mu_F)}{\pi} \, \int_{z}^{1} \frac{dy}{y} \, D_{q\to
  \gamma}\left(\frac{z}{y},\mu_F \right) \, P^{(0)}_{q\to q}(y) \, ,
\ee
where the Altarelli-Parisi splitting function  $P^{(0)}_{q\to q}(y)$ reads
\be
P^{(0)}_{q\to q}(y)=C_F \, \left[ \frac{1+y^2}{1-y} \right]_{+} \, 
\ee
and the function $P^{(0)}(z)$ is given in eq. (\ref{splittingfour}).

From the stucture of this equation it is clear that
the fragmentation functions $D_{q\to \gamma}(z_0,\mu_b)$ at a given
value of $z_0$ only depends on the initial condition $D_{q\to \gamma}(z,m_Z)$ for values
of $z$ satisfying $z \ge z_0$. This is important, because the initial condition extracted
from experiment is only known above $z>0.7$. This then means that we 
can determine $D_{q\to \gamma}(z,\mu_b)$ for values of $z \ge 0.7$
which is sufficient for our application.

We solved this equation numerically, using (\ref{fragansatz}) as
initial condition. By doing so, we performed the integration
w.r.t. $\mu_F$ using 4000 steps (at step 0 $\mu_F=m_Z$ and at step
4000 $\mu_F=\mu_b$). After each step, we fitted the
$z$-dependence to a set of 15 ``basis functions''. At the end of this
procedure, we got the fragmentation function at the low scale $\mu_b$
in a version where the $z$-dependence is given in a parametrized form.

As our application is rather sensitive to the fragmentation function
near $z=1$, we also solved (as a check) the evolution equation in
moment space which we could basically do in an analytic way. Through
this check, we are sure that the purely numerical uncertainties in our
prediction of $d\Gamma(b \to u \bar{u} d \gamma)/dE_\gamma$ are
negligible.

With the fragmentation function at the low scale
$\mu_F=\mu_b$ at hand, we can numerically evaluate, using eq. (\ref{summar}), the tree level
contributions of $P^u_1$ and $P^u_2$ to the (integrated) photon energy spectrum.
We are mostly interested in an upper limit for these contributions,
which amounts to use a small value for $\mu_0$ in eq. (\ref{fragansatz}).
Therefore we choose $\mu_0=0.02$ GeV, which is still compatible with
the range in eq. (\ref{rangetwojet}) obtained through the 2-jet cross
section at LEP. This compatibility is illustrated in Fig.~\ref{fit}.

\begin{figure}[h]
\begin{center}
\includegraphics[width=9cm]{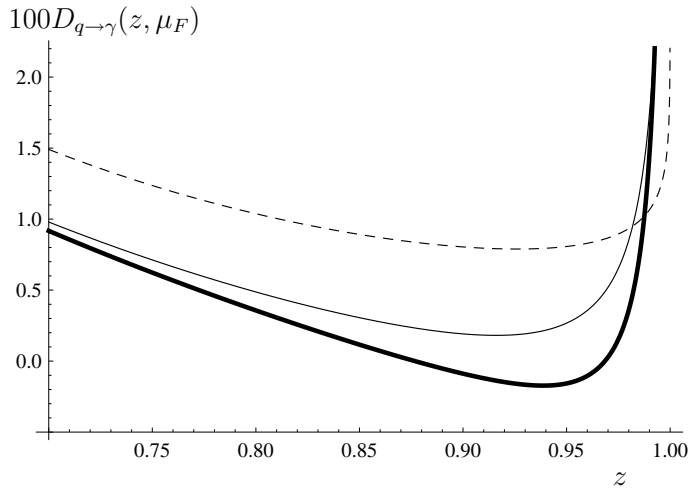}
\caption{Dependence of the fragmentation function
$D_{q\to \gamma}(z,\mu_F)$ (for $e_q=1$) on $z$ and $\mu_F$. 
The dashed line shows the fragmentation function for $\mu_F=m_Z$ as extracted
from the ALEPH data, using $\mu_0=0.02$ GeV. 
The  thick (thin) solid lines shows the corresponding fragmentation function for
$\mu_F=m_b/2$ ($\mu_F=m_b$), obtained after solving the QCD evolution
equation (\ref{llevolution})  
in leading logarithmic precision (see text).  
\label{frag}}
\end{center}
\end{figure}
The $z$-dependence of the resulting fragmentation function $D_{q \to \gamma}(z,\mu_F)$
is shown in Fig.~\ref{frag} for various values of $\mu_F$.
\subsection{Numerical results}
\label{numres}
With eq. (\ref{summar}) and the fragmentation function $D_{q \to
\gamma}(z,\mu_F)$ at the low scale $\mu_F \sim m_b$ 
we have all the ingredients to do the numerics for
$d\Gamma(b \to u \bar{u} d \gamma)/dE_\gamma$ associated with the
operators $P^u_1$ and $P^u_2$. Unless stated otherwise, we use the value
$\mu_0=0.02$ GeV in eq. (\ref{fragansatz}), because our aim is to 
give an estimate for the upper limit of this contribution.
For the other input
parameters we use $m_b=4.68$ GeV, $|\xi_u|^2=1.114 \times 10^{-5}$, 
$|\xi_t|^2=7.530 \times 10^{-5}$ and for the  
Wilson coefficients in leading logarithmic approximation 
(which are always taken at the scale 2.5 GeV in this paper) 
we use, as in ref. \cite{Kaminski:2012eb}, 
$C_1=-0.8144$, $C_2=1.0611$, $C_7=-0.3688$.

\begin{figure}[h]
\begin{center}
\includegraphics[width=9cm]{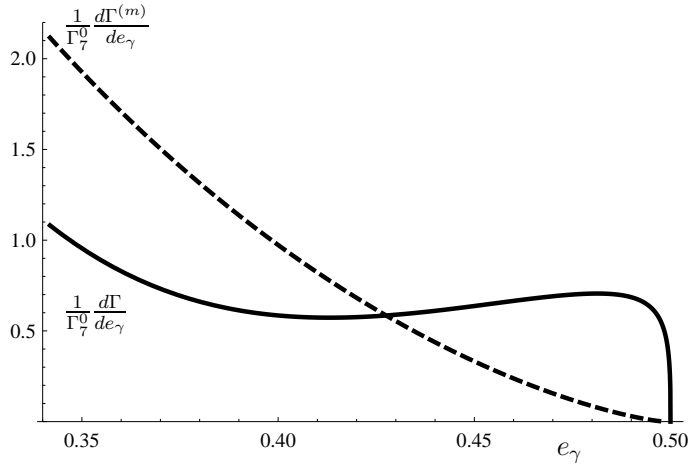}
\caption{Normalized photon energy distribution 
(\ref{specnorm}) due to $b \to u \bar{u} d \gamma$ associated 
with the operators $P^u_{1,2}$ for the two different approaches. 
Solid line: fragmentation function approach with fragmentation 
scale $\mu_F=m_b/2$;
dashed line: introducing a common constituent quark mass $m$ for the light
quarks, taking $m/m_b=1/50$. 
\label{distr}}
\end{center}
\end{figure}
In Fig.~\ref{distr} we plot the normalized photon energy spectrum
$dR_d/de_{\gamma}$  
\be
\label{specnorm}
\frac{dR_d}{de_\gamma} = \frac{1}{\Gamma_7^0}\frac{d\Gamma(b \to u \bar{u} d
  \gamma)}{de_\gamma} \,
\ee
as a function of the rescaled photon energy $e_\gamma$ 
($e_\gamma=E_\gamma/m_b$). $\Gamma_7^0$ corresponds to the total 
$b \to d \gamma$ decay width
when only taking into account the tree level contribution of the
magnetic dipole operator $P_7$, i.e,
\be
\Gamma_7^0 = \frac{G_F^2 \, m_b^5 \, |\xi_t|^2 \, \alpha \, C_7^2}{32 \, \pi^4}
\, .
\ee
The result for $dR_d/de_\gamma$ is shown by the solid line in Fig.~\ref{distr}; the dashed
line corresponds to the result when using constituent masses for the
light quarks, as will be discussed in section~\ref{secmassive}.
 
\begin{table}[h]
\begin{center}
\begin{tabular}{|c|l|l||l|l|}
\hline
$ $&$ \mu_F=m_b/2        $&$ \mu_F=m_b $&$\frac{m}{m_b}=\frac{1}{50} 
$&$ \frac{m}{m_b}=\frac{1}{10} $\\\hline
$ R_d^{cut} $&$ 0.107  $&$0.0683 $&$ 0.126 $&$ 0.0188 $\\\hline
\end{tabular}
\caption{The ratio $R_d^{cut}$ (see eq. (\ref{rdcut})) for $E_\gamma^{cut}=1.6$ GeV
 for different values of fragmentation scale $\mu_F$
and different values of the common constituent mass $m$ of the light
quarks. \label{table1}}
\end{center}
\end{table}
In Table~\ref{table1} we consider the corresponding integrated quantity $R_d^{cut}$
\be
\label{rdcut}
R_d^{cut} = \int_{e_\gamma^{cut}}^{1/2} \frac{dR_d}{de_\gamma} \, ,
\ee
for $e_\gamma^{cut}=0.342$ (which corresponds to a photon energy cut
of 1.6 GeV), using two different values for the fragmentation scale $\mu_F$.
As in Fig.~\ref{distr}, we also show in Table~\ref{table1} the
corresponding results when using a common constituent
mass $m$ for the
light quarks, as discussed in section~\ref{secmassive}.

As mentioned above, the results in Table~\ref{table1} for the
fragmentation function
approach are based on using the value $\mu_0=0.02$ GeV in
eq. (\ref{fragansatz}) and should therefore be considered as an upper
limit for $R_d^{cut}$. For the central value $\mu_0=0.14$ GeV 
(see eq. (\ref{rangetwojet})) one gets $R_d^{cut}=0.0549$ for $\mu_F=m_b/2$ and
$R_d^{cut}=0.0291$ for $\mu_F=m_b$.
 
One sees from Table~\ref{table1} that these upper limits are close
to the results when using a common constituent quark mass $m$ (with
$m/m_b=1/50$) for the light quarks.
These contributions are not very small; 
therefore it will be necessary to take them into account when deriving a NNLL 
prediction of the CP averaged branching ratio for  
$\bar{B} \to X_d \gamma$. 

\section{Result when using constituent quark masses}
\label{secmassive}
Another possibility to effectively treat the collinear regions
  connected with photon
emission from light quarks is to provide the latter with constituent masses 
\cite{Kaminski:2012eb}.

Making use of ref. \cite{Asatrian:2012tp} where useful
ingredients for computing the phase space integrals  
with massive particles in the final state are given,
we easily get the spectrum $d\Gamma^{(m)}(b \to u \bar{u} d \gamma)/dE_\gamma$  
associated with the operators $P^u_1$ and $P^u_2$. Providing all light
quarks with the same constituent mass $m$ and keeping the
$m$-dependence only in logarithmic terms, we obtain
\bea
\frac{d\Gamma^{(m)}(b \to u \bar{u} d \gamma)}{dE_\gamma} &=&
\frac{G_F^2 \, m_b^4 \, |\xi_u|^2 \, \alpha} {32\pi^4} \, \frac{\left(9 \,
  C_2^2+2 \, C_1^2 \right)}{3} \, \frac{\left(1-2 \, e_\gamma \right)}{972 \, e_\gamma} \times
\nonumber \\
&& \left[ 6 \, \left(272 \, e_\gamma^3 - 176 \, e_\gamma^2 + 44 \,
  e_\gamma - 27 \right) \, \left( 2 \ln \frac{m}{m_b} - \ln (1-2 \,
  e_\gamma) \right) \right. \nonumber \\
&& \left. \hspace{7pt} + 4316 \, e_\gamma^3 -2138 \, e_\gamma^2 +422
  \, e_\gamma -399 \right] \, .
\eea
In this formula the charge factors $e_u=2/3$ and $e_d=-1/3$ are
inserted and $e_\gamma$ stands again for the rescaled photon energy
($e_\gamma=E_\gamma/m_b $).
The numerical results of this approach can be seen in Fig.~\ref{distr}
and in Table~\ref{table1}.

\section{Summary and conclusions}

Using data from the two-jet cross section (where one of the jets is required 
to contain a photon) measured by the ALEPH experiment at LEP 
\cite{Buskulic:1995au}, the
quark-to-photon fragmentation function was extracted (with the help of
the theoretical work \cite{Glover:1993xc}) at the
fragmentation scale $\mu_F=m_Z$. 
Using this input, we determine the fragmentation function at the scale
$\mu_F \sim m_b$ 
by numerically solving the corresponding evolution equation.
Using the so-obtained  fragmentation function, we worked out the upper
limit of that
contribution to the (integrated) photon energy spectrum for $\bar{B}
\to X_d \gamma$ which stems from the tree-level transitions 
$b \to u\bar u d\gamma$ assocoated with the operators
and $P^u_{1,2}$. This upper limit is close
to the result when using a common constituent quark mass 
$m$ (with $m/m_b=1/50$) for the light quarks. 
We conclude, that these contributions are not very small and therefore 
it will be necessary to take them into account when deriving a NNLL 
prediction of the CP averaged branching ratio for  
$\bar{B} \to X_d \gamma$. Needless to say, it would be useful to have a
determination of the quark-to-photon fragmentation function
which directly uses data from $B$-meson decays. 
This would obviously lead to a more precise 
prediction of the $b \to u\bar u d\gamma$ transition.


\section*{\normalsize Acknowledgments}
This work was partially supported by the Swiss National Foundation.
H.M.A. was supported also by the AEC, the 
Volkswagen Stiftung Program No. 86426 and 
the State Committee of Science of 
Armenia Program No. 11-1c014.
We thank Aude Gehrmann-De
Ridder, Thomas Gehrmann and Massimiliano Procura for very useful discussions.

\end{document}